\documentclass[pra,showpacs,twocolumn,amsmath,amssymb,superscriptaddress,showkeys]{revtex4-1}
\usepackage{graphicx}
\usepackage{float}
\usepackage{color}
\usepackage{ulem}
\usepackage{bm}
\usepackage{hyperref}

\def\no{\noindent}

\begin{document}
\title{Bose-Einstein condensation of light in a cavity}
\author{Alex Kruchkov\\
Institute of Condensed Matter Physics, \'Ecole Polytechnique F\'ed\'erale de Lausanne (EPFL), 
\\
Station 3, CH-1015 Lausanne, Switzerland 
\\
alex.kruchkov@epfl.ch}      

\date{\today}

\begin{abstract}
The paper considers Bose-Einstein condensation (BEC) of light in a cavity with medium.
In the framework of two-level model we show the effect of gaseous medium on the critical temperature of light condensation in the system.
Transition of the system to  the state with released light condensate is illustrated in consequent stages.
Analytical expressions for a typical spatial extent of the condensed cloud of photons, as well for spectral characteristics of the condensate peak are derived. 
Energy and heat capacity of photons as functions of  temperature are obtained.
 Finally, we demonstrate that the energy of light can be accumulated
 in the BEC state.

\end{abstract}

\pacs{
67.10.Fj, 
67.85.Jk.
}

\keywords{photons in matter, thermodynamic equilibrium, electromagnetic field in a cavity, Bose-Einstein condensation of photons.}

\maketitle

\section{Introduction}

Despite the fact that Bose-Einstein distribution was first obtained for photons \cite{Bose}, 
until recently the possibility of experimental observation of Bose-Einstein condensation of light seemed to be quite absurd.
Reasons for such a common position were mainly the following obvious circumstances.
First, it is difficult to imagine a situation where photons, which are considered to be massless particles, could exist in a vacuum in the lowest energy state, i.e. with infinite wavelength.
Second, it is difficult to implement a system with the conserved number of photons for arbitrary description parameters of  a system.
In other words, to observe Bose-Einstein condensation it is necessary to ensure  the non-vanishing chemical potential of photons.
These difficulties do not arise (or at least not in such an extent) in other boson systems. 
Recent years were marked by the series of successful experimental realizations of Bose-Einstein condensation in a wide variety of many-particle systems: from ultracold gases of atoms and molecules \cite{1,2,3,4,5}  to exciton polaritons systems \cite{polariton1,polariton2}.
Therefore, despite the insuperable (at first glance) hindrances,  physicists have been again and again reverting to the tempting idea of condensing  light.

A way to get around the first constraint (lack of photon mass) was first proposed in Ref. \cite{Chiao}.
In that paper, the authors considered the behavior of photons in a cavity formed by two parallel planes with high reflectivity (for similar ideas see also Refs.\cite{Martini,Demokritov}).
Due to the finiteness of the system's longitudinal size  (transversal size is infinite in this model), the longitudinal component of the electromagnetic field inside a cavity is quantized.
This in turn leads to the appearance of an effective mass in the dispersion of photons in a cavity.
The uniqueness of a choice of the effective mass is provided by exciting the corresponding longitudinal cavity mode by external laser radiation from the red side of resonance.
We also note that although in that paper the authors had  discussed the Bose-Einstein condensation and superfluidity of photons, one cannot obtain directly through this approach the critical temperature of Bose condensation because of the logarithmic divergence of integrals in the model of two-dimensional homogeneous bosonic systems.
In addition, in the framework of Ref.\cite{Chiao}, photons occupy a vacuum cavity, without any medium, and therefore one can hardly imagine the establishment of thermodynamic equilibrium in the system that is necessary to observe Bose-Einstein condensation (BEC).

Therefore, the key issue became the question of a possible thermalization of a quasiconstant number of photons in a Fabry-Perrot cavity or a similar one.
This issue has been resolved rather recently \cite{thermalization}.
In that study, the authors were able to achieve thermodynamic equilibrium of photons in a cavity filled with a liquid organic dye, owing to the balance between the processes of an absorption and re-emission of photons by dye molecules.
A total number of photons is conserved on average, and possible small losses (for example, due to imperfections of mirrors) are compensated for by periodic weak pumping.
In a subsequent work \cite{Klaers}, the authors have produced compelling arguments in favor of the experimental observation of BEC of light.

To facilitate the localization of photons inside a microcavity in experiments Refs.\cite{thermalization,Klaers}, the authors used slightly curved mirrors. 
Due to the finite curvature of cavity walls, photons tend to be stored in the geometric center of the cavity under the effect of purely geometric pseudopotential.
In addition, the presence of this potential in photon dispersion enables us to determine the finite critical temperature of Bose condensation.
Upon reaching the condensation point,  a blurred photonic "cloud"  transforms into a bright narrow spot.
As in the case of  ultracold-gas experiments, it demonstrates visually the phenomenon of Bose condensation of light in the system under study.
Simultaneously, in the inverted space of photon wave vectors (or their energies), one can observe a narrow and high  peak, corresponding to the amount of condensed photons in the system.

The experiment described in Refs.\cite{thermalization, Klaers} is relatively simple from the viewpoint of  implementation and observation of Bose condensation in other many-particle systems (see for example, \cite{1,2,3,4,5}).
However, the result is so intriguing that it gives a stimulus to a substantial number of theoretical studies on the subject \cite{Kruchkov,Sobyanin1,Sobyanin2,Snoke,Kirton,Zhang1,Zhang2,Zhang3,Leeuw}.
Despite the considerable progress in this area, the phenomenon of BEC of photons needs further study.

In particular, it would be interesting to consider in full the problem of determining the effect of medium inside cavity on the parameters of Bose condensation of photons in equilibrium with the medium.
In other words, it is unclear in what way the medium that is so needed for thermal equilibrium to be achieved, is included in the final equation for determining the critical parameters of the system in Refs.\cite{thermalization,Klaers}.
This question was to some extent studied in Ref.\cite{Kruchkov}, but the system considered therein is spatially isotropic and homogeneous, which  is rather far from current experimental conditions \cite{thermalization, Klaers}; and also in Ref.\cite{Sobyanin2} from the first principles, however no direct answer in explicit form was obtained. So in the present paper we propose the plane answer:  The medium can redefine photon's effective mass, therefore changing the critical temperature and other parameters.

One more rather important issue is the consequent derivation of statistical properties, and primarily the form of the effectively two-dimensional distribution function of photons in a cavity.
Recall that the authors of Refs.\cite{thermalization, Klaers}  used for numerical estimates the form which is valid for the truly two-dimensional Bose condensates.
In our opinion, to obtain using this approach the correct expressions for Bose condensation parameters in a consistent and non-contradictory way is rather difficult.
We give an expression for the correctly normalized two-dimensional distribution function, which leads to the proper definition of the critical temperature.

In this paper we mainly pay attention to the  above-mentioned issues.
Sec. II describes one of the ways to build a statistical description of photons in a cavity with medium.
Special attention is given to a favorable case of thermodynamic equilibrium.
At this stage, in-cavity medium characteristics are introduced into the distribution function of photons. 
This is to some extent a generalization of the approach used previously by the authors of Ref.\cite{Kruchkov}.
In Sec. III, we demonstrate the evolution of the system to the phase transition point; we derive consequently the statistical and thermodynamic characteristics of the photonic component of the system. Finally, we show explicitly that  light energy can be accumulated in Bose-condensed photons.

\section{Statistical description of photons in a cavity in the presence of gaseous medium}
\label{2}

On the kinetic stage of evolution, properties of photons, which can be absorbed and re-emitted by medium inside a cavity, can be described with the kinetic equation for a distribution function of photons as (quasi)particles.
However, the use of only kinetic equation is insufficient. From the general assumption it is clear that the behavior of the distribution function of photons is affected by a dispersion law of photons inside a cavity, which should include the consideration of boundary conditions in a self-consistent way.

The dispersion relation for photons in a cavity with the presence of medium, generally speaking, should be determined on the basis of Maxwell's equations supplemented by the boundary conditions on the surfaces of mirrors. In the case of the narrow cavity,  confined  on both sides by two coaxial sections of an ideal mirror, the longitudinal modes of the electromagnetic field in the cavity are discrete. Thus for the longitudinal wavenumber $k_{z}$ in a case of ideally reflecting boundary conditions one obtains

\begin{equation}\begin{split}\begin{gathered}
\label{kz}
{{k}_{z}}=\frac{\pi q}{l\left( r \right)},
\end{gathered}\end{split}\end{equation}

\begin{figure}[t]
\center{\includegraphics[width=0.8\columnwidth]{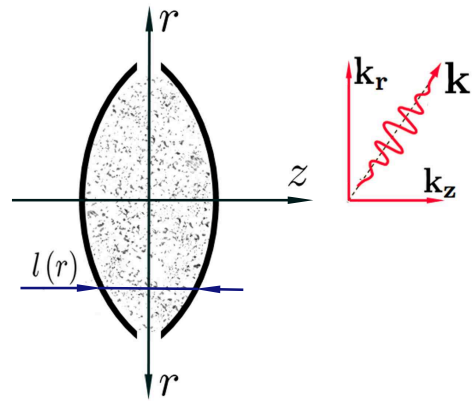}}
\caption{(Color online)
A cavity with  thermostatic gaseous medium bordered by two spherically curved mirrors with a distance $l(r)$ between them.
Due to the symmetry of the system, the state space of photon inside the cavity is characterized only by longitudinal $k_{z}$ and transverse $k_{r}$ wave numbers. }
\label{scheme}
\end{figure}

\no
where $q$ is a longitudinal mode number, and $l\left(r \right)$ is a cavity width depending on transversal radius $r$ (see Fig.~\ref{scheme}). Therefore we actually proceed to the cylindrical system of coordinates, where, due to the symmetry of the system under study, photon modes are fully described by two components of wave vector, $k_{z}$ and $k_{r}$ respectively. In terms of these components, photon energy can be determined as:

\begin{equation}\begin{split}\begin{gathered}
\label{energy}
{{\varepsilon }^{*}}=\hbar \tilde{c}k=\hbar \tilde{c}\sqrt{k_{z}^{2}+k_{r}^{2}},
\end{gathered}\end{split}\end{equation}

\no
where $\tilde{c}$ is the speed of light in medium. 
For the sake of simplicity in further expressions throughout this paper the tilde sign is omitted.

As we have already mentioned, the longitudinal wavenumber is the set of discrete constants \eqref{kz}, and currently one can thermalize only one of these modes in experiment (see Refs. \cite{thermalization, Klaers}). 
Therefore, one can treat formula \eqref{energy} as the expression for relativistic energy of quasiparticle with some effective mass that depends on $k_{z}$. 
Taking into account a spherical curvature of mirrors to determine the distance between them $l\left(r \right)$, and considering $l\left(r \right)$ much smaller than a mirror size, one can expand Eq.\eqref{energy} as follows (see also \textit{paraxial approximation} in Refs. \cite{thermalization, Klaers,Sobyanin2}):

\begin{equation}\begin{split}\begin{gathered}
\label{E*}
{{\varepsilon }^{*}}={{m}^{*}}{{c}^{2}}+\frac{{{\hbar }^{2}}k_{r}^{2}}{2{{m}^{*}}}+\frac{1}{2}{{m}^{*}}{{\Omega }^{2}}{{r}^{2}},
\end{gathered}\end{split}\end{equation}

\no
where we have introduced the photon effective mass

\begin{equation}\begin{split}\begin{gathered}
\label{m*}
{{m}^{*}}\equiv \frac{\hbar {{k}_{z}}\left( 0 \right)}{c},
\end{gathered}\end{split}\end{equation}

\no
as well as the effective frequency $\Omega$ of the harmonic pseudo-potential:

\begin{equation}\begin{split}\begin{gathered}
\label{Omega}
{{\Omega }^{2}}\equiv \frac{2{{c}^{2}}}{{{l}_{0}}R}.
\end{gathered}\end{split}\end{equation}

\no
In the last expression, $R$ is the curvature radius of mirrors, and ${l}_{0} \equiv l(0)$ is the width of the cavity on the symmetry axis.

We emphasize here that, according to Eq.\eqref{E*}, photons behave themselves as if they were two-dimensional particles with a mass $m^{*}$, moving with kinetic energy ${\hbar }^{2} k_{r}^{2}/2m^{*}$ in external harmonic potential $\Omega$. It leads us to a great analogy with the phenomenon of Bose condensation in trapped ultracold gases. In terms of photons, Bose condensation means the macroscopical occupation of the lowest energy level, i.e. with $k_{r}=0$. Therefore, conducting experiments for spectral density of radiation inside cavity, one would measure the peak, corresponding to the frequency

\begin{equation}\begin{split}\begin{gathered}
\label{omega}
 \omega_{0}=\frac{m^{*} c^{2}}{\hbar}. 
\end{gathered}\end{split}\end{equation}

\no
In recent experiments  \cite{thermalization, Klaers} photons have been condensed to the state with the frequency  $ \omega_{0}=3.2 \cdot 10^{15} \, s^{-1}$ (that corresponds to $\lambda_{0}=584 \, nm$).

We define the distribution function of photons  $f_{\gamma} \left( \mathbf{r},\mathbf{k}, t \right) $ as the total amount of photons (both free particles and absorbed during atomic transitions quanta of field), localized at time $t$ in the unit  phase volume $\left\{ \mathbf{r},\mathbf{k} \right\}$.
The behavior of the distribution function of photons in cavity on the kinetic stage of evolution obeys the kinetic equation 

\begin{equation}\begin{split}\begin{gathered}
\label{kinetic equation}
\frac{d}{dt}f_{\gamma} \left( \mathbf{r},\mathbf{k}, t  \right)={{I}_{coll}} \,  \sigma(\mathbf{r}).
\end{gathered}\end{split}\end{equation}

\no
One should consider the  derivative in the left side of the equation \eqref{kinetic equation} as a  so-called substantial derivative (see Refs.\cite{Landau,Akhiezer}).
Function $\sigma(\mathbf{r})$ takes into account the finite size of the system. In a simplified case one can consider that  $\sigma(\mathbf{r})=1$ if $\mathbf{r}$ points inside the cavity with medium, and  $\sigma(\mathbf{r})=\infty$ in other cases.
A rate of gaining thermodynamic equilibrium between photons and medium, if it is possible for given parameters of system, is determined by collision integral $I_{coll}$, introduced into the right-hand side of Eq.\eqref{kinetic equation}.
The collision integral $ I_ {coll} $ must take into account not only those processes that lead to the establishment of thermodynamic equilibrium, but also the possibility of absorption and re-emission of photons by medium even in equilibrium state of system.
In other words, the balance between the processes of creation and annihilation of photons should be also included in the distribution function  $f_{\gamma} \left( \mathbf{r},\mathbf{k}, t \right) $.

From general considerations, however, it is clear that the limiting case of thermodynamic equilibrium should be described in a  much simpler way than processes that result this state.
For time periods to be much more than relaxation time, the correlations between free and bound photons  disappear almost entirely.
Thereafter, one naturally considers the distribution function of photons, thermalized with medium inside cavity, to be written as

\begin{equation}\begin{split}\begin{gathered}
\label{distr gen}
f_{\gamma}  \left( \mathbf{r},\mathbf{k}  \right)=f\left( \mathbf{r},\mathbf{k}  \right)+\tilde{f}\left( \mathbf{r},\mathbf{k}  \right),
\end{gathered}\end{split}\end{equation}

\no
where $f\left( \mathbf{r},\mathbf{k}  \right)$ corresponds to the equilibrium Bose distribution function of free photons with dispersion law \eqref{E*} and the chemical potential of photons $\mu^{*}$,

\begin{equation}\begin{split}\begin{gathered}
\label{distr free}
f\left( \mathbf{r},\mathbf{k}  \right) =
g^{*}
{{\left\{ \exp \left( \frac{{{\varepsilon }^{*}}-{{\mu }^{*}}}{T} \right)-1 \right\}}^{-1}},
\end{gathered}\end{split}\end{equation}

\no
where  $T$ is the temperature of the system in energy units,
and
the quantity $g^{*}$  takes into account the possible degeneracy in  photon energy (see  Ref.\cite{Kruchkov}).
The degeneracy of the photon modes, which in principle could depend on the photon energy $\varepsilon^{*}$, throughout the present paper can be considered as some effective degeneracy $g^{*} \approx 2$ (see the discussion in the Conclusion section). 
Distribution function \eqref{distr free} should be normalized on the total amount of free photons $N$ in the system under study,

\begin{equation}\begin{split}\begin{gathered}
\label{Nfree}
\iint{
\frac{d\mathbf{k}d\mathbf{r}}{{{\left( 2\pi  \right)}^{3}}}
\,
f\left( \mathbf{r},\mathbf{k} \right)
}
=N.
\end{gathered}\end{split}\end{equation}

\no
Correspondingly, the physical quantity $\tilde{f}\left( \mathbf{r},\mathbf{k}  \right)$ demonstrates the portion of photons that are bounded with structural units of gaseous medium (atoms or molecules) to create the excited states of these units. Similarly, this distribution function  is normalized to the total amount of bounded photons $\tilde{N}$ in the system under study,

\begin{equation}\begin{split}\begin{gathered}
\label{Nbound}
\iint{
\frac{d\mathbf{k}d\mathbf{r}}{{{\left( 2\pi  \right)}^{3}}}
\,
\tilde{f}\left( \mathbf{r},\mathbf{k} \right)
}
=\tilde{N}.
\end{gathered}\end{split}\end{equation}

\no
As a consequence of Eq.\eqref{distr gen}, the total number $N_{\gamma}$ of pumped photons conserves:

\begin{equation}\begin{split}\begin{gathered}
\label{Ngamma}
N_{\gamma}=N+\tilde{N}.
\end{gathered}\end{split}\end{equation}

\no
We emphasize here that only the mean occupation numbers of photons in the system have an  effect on the expectation values of measurable quantities, and therefore throughout the paper we do not take into account their fluctuations. For the effect of fluctuations in light BEC see e.g. Ref. \cite{Sobyanin2}.

We emphasize that the quantities $f\left( \mathbf{r},\mathbf{k}  \right)$ and  $\tilde{f}\left( \mathbf{r},\mathbf{k}  \right)$ in expression \eqref{distr gen} and therefore in kinetic equation \eqref{kinetic equation} are purely three-dimensional. Reduction to the effectively two-dimensional problem should be introduced by consequent procedure, taking into account the geometry of the cavity. 
The formal way to do it is to integrate three-dimensional distribution function with respect to $z$ and then sum for all possible values of $k_{z}$. For example, a two-dimensional distribution function of free photons is obtained following this procedure and taking into account Eqs.\eqref{distr free},\eqref{E*} and also that only one longitudinal mode $q$ survives (see the Appendix):

\begin{equation}\begin{split}\begin{gathered}
\label{distr 2D}
{{\left. f\left( r,{{k}_{r}} \right) \right|}_{2D}}
=\sum\limits_{{{k}_{z}}}{\int\limits_{z}{\frac{dz}{2\pi }}}
f\left( \mathbf{r},\mathbf{k}  \right)
\\
=\frac{g^{*}q}{2}
{{\left\{ \exp \left( \frac{{{\hbar }^{2}}k_{r}^{2}}{2{{m}^{*}}T}+\frac{{{m}^{*}}{{\Omega }^{2}}{{r}^{2}}}{2T}
+\frac{m^{*}c^2-\mu^{*}}{T}
 \right)-1 \right\}}^{-1}}.
\end{gathered}\end{split}\end{equation}

As one can see from Eq.\eqref{distr 2D}, the two-dimensional  distribution function of free photons $ f\left( r,{{k}_{r}} \right)$ depends on longitudinal mode number $q$. Consequently, the critical number of photons to observe Bose condensation also depends on $q$ (see Sec. III).

Note also the following circumstance. 
In the large range of gaseous medium parameters, the amount of bounded photons can be relatively weak, $\tilde{f}\left( \mathbf{r},\mathbf{k}  \right) \ll f\left( \mathbf{r},\mathbf{k}  \right)$. Therefore, from a mathematical point of view  one can treat $\tilde{f}\left( \mathbf{r},\mathbf{k}  \right)$ as the a perturbation of  $f\left( \mathbf{r},\mathbf{k}  \right)$ in Eq.\eqref{distr gen}.
Taking into account Eqs.\eqref{distr free},\eqref{E*} one may state that  the perturbation parameter is proportional to the photon effective mass \eqref{m*}  for other parameters  of the system (temperature, chemical potential of photons etc.) to be constant.
In this sense, one can contend that interactions between photons and matter redefine the photon effective mass.

\section{Statistical and Thermodynamical properties of Bose-Einstein condensation of photons in a cavity}

To define a transition temperature for the Bose-Einstein condensate of photons in matter, one needs to introduce a condition for the chemical potential of photons similar to those ones in the theory of atomic BEC
(for details see Ref.\cite{Kruchkov}):

\begin{equation}\begin{split}\begin{gathered}
\label{mu*}
\mu ^{*} \left(T\le T_{c} \right)=m^{*} c^{2}.
\end{gathered}\end{split}\end{equation}

\no
This condition ensures the distribution function of free photons \eqref{distr free} to be positive.
Before determining critical parameters, we examine the process of gaining the phase transition.
We introduce a two-dimensional density of free photons in a cavity normalized on the total number of free photons $N$:

\begin{equation}\begin{split}\begin{gathered}
\label{2D density}
n\left( r \right)
=
\frac{1}{(2 \pi)^3}
\sum\limits_{{{k}_{z}}}{\int\limits_{z}{dz}\int\limits_{{{\mathbf{k}}_{r}}}{{d {{\mathbf{k}}_{r}}} {f \left( \mathbf{r},\mathbf{k}  \right)}}},
\end{gathered}\end{split}\end{equation}

\begin{figure*}
\center{\includegraphics[width=0.9\textwidth]{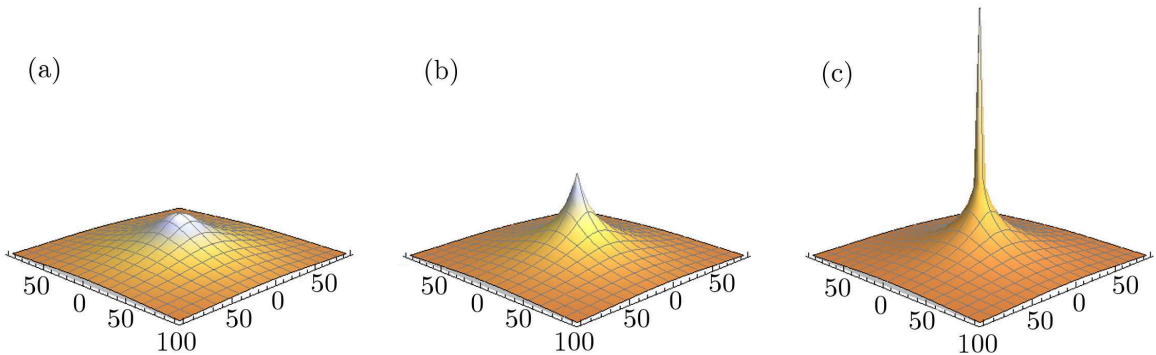}}
\caption{(Color online)
Consequent stages of evolution of the system towards the critical point of Bose-Einstein condensation of photons: (a) and (b) - before the condensation starts, (c) - at the BEC phase transition point.
The height of every three-dimensional plot corresponds to the spatial density of free photons $n(r)$ (in arbitrary units) plotted as a function of Cartesian coordinates $\left \{x,y \right\}$ (in $\mu m$) for different chemical potentials of photons $\mu^{*}$. Note that in panel  (c) the height of the peak, $n(0)$, is actually infinite. Naturally, condensation of light starts at the geometrical center of the cavity, $r=0$ . }
\label{density}
\end{figure*}

\no
Calculating the integrals in the expression \eqref{2D density} and using definition \eqref{distr free}, one can easily manifest an explicit expression for the upside of the critical point in the system under study:

\begin{equation}\begin{split}\begin{gathered}
\label{density profile}
n\left( r \right)=\frac{{{g}^{*}}q}{4\pi }\frac{{{m}^{*}}T}{{{\hbar }^{2}}}
\\
\times
\ln {{\left\{ 1-\exp \left[ \frac{1}{T}\left( {{\mu }^{*}}-m^{*} c^{2} -\frac{1}{2}{{m}^{*}}{{\Omega }^{2}}{{r}^{2}} \right) \right] \right\}}^{-1}}.
\end{gathered}\end{split}\end{equation}

\no
According to the definition of the critical point \eqref{mu*} and the density of free photons \eqref{distr free}, the chemical potential of photons cannot overflow the value of $m^{*} c^2$. Consequently, one can plot a spatial distribution of light intensity in the cavity, using the expression  \eqref{density profile} 
 for the photon density profile $n\left( r \right)$ with different chemical potential $\mu^{*}$,
and it will correspond to the different temperatures above the critical one.

Figure~\ref{density} shows the three-dimensional plots showing the spatial dependence for the density  of free photons $n\left( r \right)$ for three different chemical potentials:
In panels (a) and (b) - before the criticality (where $\mu^{*} < m^{*}c^2$)  and in the case (c) - at the critical point (where $\mu^{*}=m^{*}c^2$). Figure 2 was plotted for parameters of the system, similar to parameters of experimental setup \cite{Klaers}: the longitudinal mode number $q=7$,  and the effective potential $\Omega \approx 3 \cdot 10^{11} \, s^{-1}$, photon's effective mass $m^{*}$ in order of $10^{-35} \, kg$ and room temperature $T=300 \, K \approx 2.6 \cdot 10^{-2} \, eV$. Note that the peak on  plot (c), where the light condensation starts, is actually infinite but cut due to our restricted visualization possibilities. For these reasons we do not introduce the vertical scale for $n(r)$.

We can define luminosity of light radiation inside cavity as an energy transferred through the unit surface  \footnote{In radiometry this quantity is often called  a photometric exposure.} . This quantity can be calculated as the function of radial coordinate by summing all the energies of thermalized photons inside cavity with corresponding weight given by the Bose-Einstein distribution function:

\begin{equation}\begin{split}\begin{gathered}
\label{luminosity}
I\left( r \right)=\frac{1}{{{\left( 2\pi  \right)}^{3}}}\sum\limits_{{{k}_{z}}}{\int\limits_{z}{dz}}\int\limits_{0}^{\infty }
\frac{{{g^{*} \, \varepsilon }^{*}}\left( {{k}_{r}},r \right)  {d\left( \pi k_{r}^{2} \right)}}
{\exp \left[ \frac{1}{T}\left( {{\varepsilon }^{*}}\left( {{k}_{r}},r \right)-{{\mu }^{*}} \right) \right]-1},
\end{gathered}\end{split}\end{equation}

\no
where $\varepsilon^* \left( {{k}_{r}},r \right)$ is given by the expression  \eqref{E*}. Taking now into account expression \eqref{density profile}, and introducing a dimensionless quantity 
$\xi ={{\hbar }^{2}}k_{r}^{2}/2{{m}^{*}}T $
, one can rewrite \eqref{luminosity} in a more convenient form:

\begin{equation}\begin{split}\begin{gathered}
\label{luminosity2}
I\left( r \right)=\left(m^{*} c^{2}+\frac{1}{2}{{m}^{*}}{{\Omega }^{2}}{{r}^{2}} \right)n\left( r \right)
\\
+\frac{{{g}^{*}}q}{4\pi }\frac{{{m}^{*}}{{T}^{2}}}{{{\hbar }^{2}}}\int\limits_{0}^{\infty }{\frac{\xi \, d\xi }{{{e}^{\xi +u\left( r \right)}}-1}},
\end{gathered}\end{split}\end{equation}

\no
where $u(r)$ is also a dimensionless quantity that refers to "potential" energy:

\begin{equation}\begin{split}\begin{gathered}
u\left( r \right)=\frac{{{m}^{*}}{{\Omega }^{2}}}{2T}{{r}^{2}}+\frac{{{m}^{*}}{{c}^{2}}-{{\mu }^{*}}}{T}.
\end{gathered}\end{split}\end{equation}

\no
Unfortunately, the improper integral in \eqref{luminosity2} can be expressed only in terms of special functions. However, one can always calculate it numerically in order to compare with experimental observations.

\begin{figure}[b]
\center{\includegraphics[width=\columnwidth]{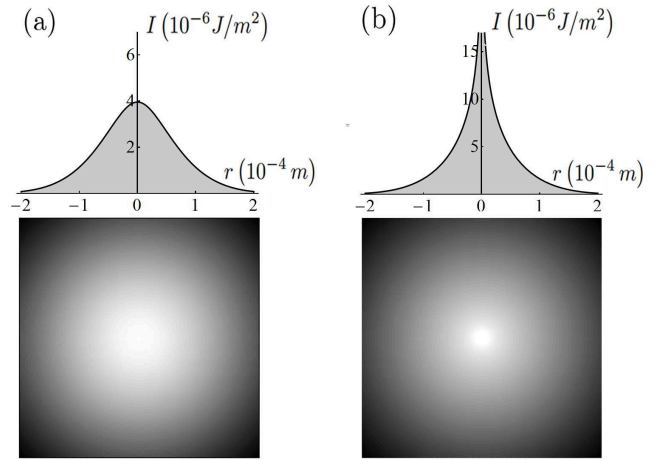}}
\caption{
Spatial distribution of light luminosity, defined by expression \eqref{luminosity}, inside a cavity: (a) - before the Bose condensation; (b) - as Bose condensation of photons starts.  One can detect BEC state by observation the bright spot in the geometrical center of a cavity.}
\label{intensity}
\end{figure}

Figure~\ref{intensity} shows the spatial distribution of the light luminosty $I(r)$ for two cases: (a) -  before the criticality ($\mu^{*} < m^{*} c^{2}$);  (b) at the critical point ($\mu^{*} = m^{*} c^{2}$).  
These figures were plotted for the parameters of the system, similar to the parameters of experimental  setup \cite{Klaers}, which were mentioned above. 
One can easily see the appearance of the bright spot in the center of a cavity at the BEC transition point  (the position of the spot  corresponds to the minimum of "potential" energy).
This completely coincides with the results of visual part of experimental observation of Bose-Einstein condensation of photons in a microcavity \cite{Klaers}, where the same bright spot is observed as the system evolves to the state with Bose-Einstein-condensed light.

So that is the reason why we need the curved mirrors: we can observe Bose-condensed photons visually.

We  calculate the critical total number of photons $N_{c}$  necessary to be pumped into a cavity to ensure the phase transition. According to the definition of the critical point \eqref{mu*}, the critical number of photons is determined by the expression:

\begin{equation}\begin{split}\begin{gathered}
\label{Ncr}
N_{c}=
{{\left. 
\iint{
\frac{d\mathbf{k}d\mathbf{r}}{{{\left( 2\pi  \right)}^{3}}}
\,
f_{\gamma}\left( \mathbf{r},\mathbf{k} \right)
}
 \right|}_{\mu^{*}=m^{*} c^2 }}.
\end{gathered}\end{split}\end{equation}

\no
Performing the integration and taking into account Eqs. \eqref{distr gen} - \eqref{Nbound}, for the cavity with photonic modes \eqref{E*} one obtains:

\begin{equation}\begin{split}\begin{gathered}
\label{Nc}
N_{c}={{g}^{*}}\frac{{{\pi }^{2}}}{12}q{{\left( \frac{{{T}_{c}}}{\hbar \Omega } \right)}^{2}}
+\tilde{N} \left( T_{c} \right),
\end{gathered}\end{split}\end{equation}

\no
where $T_{c}$ is the critical temperature of Bose-Einstein condensation in the system consisting of $N_{c}$ photons in total, among which $\tilde{N}  \left( T_{c} \right) $ are bounded with atoms at current temperature. In general, Eq. \eqref{Nc} is an intricated equation that should be solved in a self-consistent way taking into account, for example, the kinetic equation for distribution function. However,  for the sake of simplification we use the expression for the amount of photons bounded with non-interacting two-level atomic gas in the system with chemical potential of photons to be conserved (see Ref. \cite{Kruchkov}):

\begin{equation}\begin{split}\begin{gathered}
\label{Ncpld}
\frac
{\tilde{N  } \left( T \right)}
{N_{a}}
= \left( 1+\frac{{{g}_{{{\alpha }_{1}}}}}{{{g}_{{{\alpha }_{2}}}}}{{\operatorname{e}}^{\Delta /T}} \right)^{-1},
\end{gathered}\end{split}\end{equation}

\no
where $N_{a}$ is the number of two-level atoms,  $g_{\alpha_{1}}$,  $g_{\alpha_{2}}$ are degeneracies of the ground and excited atomic levels with energies $\varepsilon_{\alpha_{1}}$, $\varepsilon_{\alpha_{2}}$ correspondingly, and  $\Delta$ is  resonance detuning,

\begin{equation}\begin{split}\begin{gathered}
\label{detuning}
\Delta ={{\varepsilon }_{{{\alpha }_{2}}}}-{{\varepsilon }_{{{\alpha }_{1}}}}-\hbar {{\omega }_{0}},
\end{gathered}\end{split}\end{equation}

\no
with $\omega_{0}$ given by formula \eqref{omega}.
Therefore, substituting \eqref{Ncpld} into \eqref{Nc}, one obtains the expression to determine the necessary amount of photons, which were pumped into the system  to observe BEC:

\begin{equation}\begin{split}\begin{gathered}
\label{critical number}
N_{c}={{g}^{*}}\frac{{{\pi }^{2}}}{12}q{{\left( \frac{{{T}_{c}}}{\hbar \Omega } \right)}^{2}}
+N_{a} \left( 1+\frac{{{g}_{{{\alpha }_{1}}}}}{{{g}_{{{\alpha }_{2}}}}}{{\operatorname{e}}^{\Delta /T_{c}}} \right)^{-1}.
\end{gathered}\end{split}\end{equation}

Now we can estimate an effective number of photons to start Bose condensation inside cavity. For photons with longitudinal mode $q=7$ thermalized at room temperature $T=300 \, K$ inside the cavity with similar parameters as in Ref.\cite{Klaers}, we need around $N_{c} \approx 3 \cdot 10^5$ particles.

The number of photons in BEC state beneath the critical point  ($T<T_{c}$)  is given by the expression

\begin{equation}\begin{split}\begin{gathered}
{{N}_{{{\mathbf{k}}_{r}}=0}}(T)
=
N_{\gamma}-
\frac{{{\pi }^{2}}}{12}{{g}^{*}}q{{\left( \frac{T}{\hbar \Omega } \right)}^{2}}
\\
- N_{a} \left( 1+\frac{{{g}_{{{\alpha }_{1}}}}}{{{g}_{{{\alpha }_{2}}}}}{{\operatorname{e}}^{\Delta /T}} \right)^{-1},
\end{gathered}\end{split}\end{equation}

\no
where $N_{\gamma}$ is a total amount of photons pumped in the system [see \eqref{Ngamma}].
In the limiting case of zero temperature the number of condensed photons is

\begin{equation}\begin{split}\begin{gathered}
\label{N0}
{{N}_{{{\mathbf{k}}_{r}}=0}}\left( T=0 \right)= 
\begin{cases}
   N_{\gamma},& \text{if} \ \  \Delta >0,
\\
    N_{\gamma}-N_{a}, & \text{if} \ \ \Delta <0.
\end{cases}
\end{gathered}\end{split}\end{equation}

To facilitate experimental implementation of photonic BEC, it is desirable to make favorable conditions for increasing the number of condensed photons. According to expression \eqref{N0} one should choose the case of $ \Delta >0$, or, in other words, the red side of resonance  $\hbar \omega_{0} < \varepsilon_{\alpha_{2}} - \varepsilon_{\alpha_{1}}$. It is possible to do it because one can manage the geometry of a cavity with photons and therefore take $ \omega_{0}$ under the control.

\begin{figure}[t]
\center{\includegraphics[width=0.95 \columnwidth]{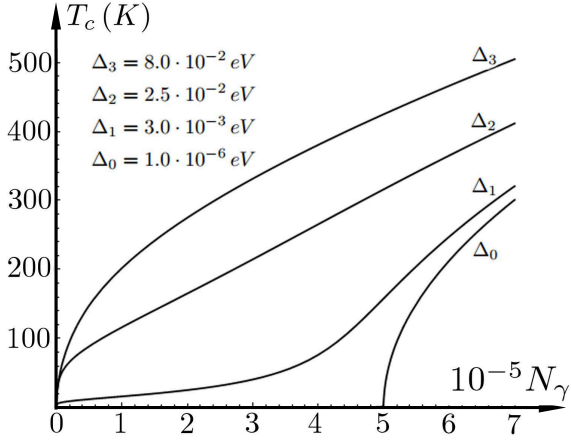}}
\caption{
Critical temperature $T_c$ as a function of the total number of photons $N_{\gamma}$ in the system, plotted according to Eq.\eqref{critical number} for different values of detuning parameter $\Delta$. The phenomenon of BEC of light demonstrates different regimes of condensation.
The resonant regime ($\Delta_{0} \approx 0$) is desirable from the viewpoint of photon gas thermalization.
}
\label{regimes}
\end{figure}

It is instructive to show how the critical temperature $T_{c}$  of light condensation depends on the total number $N_{\gamma}$ of photons  in the system, and to compare this dependence with traditional fractional-power laws in alkali atoms BEC experiments (see e.g. Ref.\cite{Pethick Smith}). Unfortunately, the equation \eqref{critical number}, which also determines the functional dependence $T_{c} \left( N_{\gamma} \equiv N_{c} \right)$, cannot be solved analitically in elementary functions of $N_{\gamma}$. Therefore, the only possibility is to solve it numerically, for the system parameters to be close to experimental ones. However, the behavior of a solution depends crucially on the detuning parameter $\Delta$, given by Eq.\eqref{detuning}, and reveals different condensation regimes. The existence of such regimes in three-dimensional systems was first studied in Ref.\cite{Kruchkov}. Figure~\ref{regimes} shows a few of these regimes that are possible for a constant number $N_a$ of gaseous medium's structural units.  Plotting Fig.~\ref{regimes}, we took into consideration that $0 < \Delta \ll \hbar \omega_{0}$: It is desirable, as we mentioned above, to have the positive detuning, but on the other hand this detuning should be rather small for thermalization of light on the gaseous medium to be possible. 
The traditional fractional-power law regime (a curve labeled as $\Delta_{3}$ on Fig.~\ref{regimes} ) is thereby only one of few possible regimes of condensation of photons (see also Ref.\cite{Kruchkov}).

We now put into consideration the spectral density of free photons $\nu_{\mathbf{k}_{r}}$ in the phase space of transversal wavevectors $\mathbf{k}_{r}$, defined by a   normalization condition:

\begin{equation}\begin{split}\begin{gathered}
\label{spectral density}
\int{{{\nu }_{{{\mathbf{k}}_{r}}}}d{{\mathbf{k}}_{r}}}=N,
\end{gathered}\end{split}\end{equation}

\noindent
where $N$ is the total number of free photons in the system that are in thermodynamical equilibrium with the components of the system. In the case of two-level atomic gas, as we have already mentioned, this quantity can be expressed explicitly:

\begin{equation}\begin{split}\begin{gathered}
\label{N*}
{{N}}
=
 N_{\gamma} - N_{a} {{\left( 1+\frac{{{g}_{{{\alpha }_{1}}}}}{{{g}_{{{\alpha }_{2}}}}}{{\operatorname{e}}^{\Delta /T}} \right)}^{-1}}.
\end{gathered}\end{split}\end{equation}

It is convenient to give the spectral density of free photons ${{\nu }_{{{\mathbf{k}}_{r}}}}$ beneath the phase transition point  in the form

\begin{equation}\begin{split}\begin{gathered}
\label{spectral density below}
{{\nu }_{{{\mathbf{k}}_{r}}}}=
{{N}_{{{\mathbf{k}}_{r}}=0}} \
\delta \left( {{\mathbf{k}}_{r}} \right)+
{{\nu }_{{{\mathbf{k}}_{r}}\ne 0}}.
\end{gathered}\end{split}\end{equation}

\noindent 
where  $\delta \left( {{\mathbf{k}}_{r}} \right)$ is the two-dimensional (2D) Dirac delta-function. The quantity ${{\nu }_{{{\mathbf{k}}_{r}}\ne 0}}$ describes the spectral density of photons with non-zero transverse momenta $\mathbf{k}_{r}$, obtained from distribution function  $f \left( \mathbf{r}, \mathbf{k}  \right)$:

\begin{equation}\begin{split}\begin{gathered}
\label{spectral density of non-condensed}
{{\nu }_{{{\mathbf{k}}_{r}}\ne 0}}
=
g^{*} 
\sum\limits_{{{k}_{z}}}{\int\limits_{V}{
\frac{dV}{(2 \pi)^{3}}
{{\left\{ \exp \left( \frac{{{\hbar }^{2}}k_{r}^{2}}{2{{m}^{*}}T}+\frac{{{m}^{*}}{{\Omega }^{2}}{{r}^{2}}}{2T} \right)-1 \right\}}^{-1}} 
}
 }
\\
=\frac{{{g}^{*}}q}{4 \pi}\frac{T}{{{m}^{*}}{{\Omega }^{2}}} \, \ln \left[ 1-\exp \left( -\frac{{{\hbar }^{2}}k_{r}^{2}}{2{{m}^{*}}T} \right) \right]^{-1}.
\end{gathered}\end{split}\end{equation}

\no
For the convenience of subsequent discussion, we introduce also  the quantity  ${{N}_{{{\mathbf{k}}_{r}}\ne 0}}$, showing the amount of non-condensed particles below the BEC transition point:

\begin{equation}\begin{split}\begin{gathered}
\label{above cond particles}
{{N}_{{{\mathbf{k}}_{r}}\ne 0}}
=
 \underset{\delta \to +0}{\mathop{\lim }}\,\int\limits_{\delta }^{\infty }
2\pi {{{k}_{r}} \, {{\nu }_{{{\mathbf{k}}_{r}}\ne 0}} \, d{{k}_{r}}}
\\
=
\frac{\pi^{2}}{12}
{{g}^{*}}q{{\left( \frac{T}{\hbar \Omega } \right)}^{2}},
\end{gathered}\end{split}\end{equation}

\no
so the total number of free photons in system under study is given as the sum of the condensed and non-condensed contributions, $N={{N}_{{{\mathbf{k}}_{r}}= 0}}+{{N}_{{{\mathbf{k}}_{r}}\ne 0}}$.

In the case when all the particles are Bose-condensed, the spectral density of photons ${{\nu }_{{{\mathbf{k}}_{r}}}}$, according to Eq.\eqref{spectral density below}, is a delta-peaked function.
Of course, in the real experiment, when one measures the spectrum of photons in a cavity, the spectral width is limited due to the finite resolution of a measuring device. Recall that the phenomenon of Bose-Einstein condensation of photons in a microcavity is recognised in essence only due to the appearance of a sharp peak in a vicinity of the cut-off frequency $\omega_{0}$  [see Ref. \cite{Klaers}]. But even before the resolution limit is gained, the peak has a finite width "blurred" due to the thermal fluctuations of non-condensed particles, and consequently the spectral width of measured signal, generally speaking, will depend on the temperature of the system.
The thermal blur of the Bose-condensed signal  can be characterized by the quantity

\begin{equation}\begin{split}\begin{gathered}
\label{spectral width def}
\hbar \left\langle \omega -{{\omega }_{0}} \right\rangle \simeq \frac{{{\hbar }^{2}}}{2{{m}^{*}}}\left\langle k_{r}^{2} \right\rangle.
\end{gathered}\end{split}\end{equation}

The mean value of the squared transverse wavevector $\left\langle k_{r}^{2} \right\rangle$ can be found with the help of averaging the quantity $k_{r}^{2}$  over the state space with the photon spectral function ${{\nu }_{{{\mathbf{k}}_{r}}}}$:

\begin{equation}\begin{split}\begin{gathered}
\label{mean k^2 def}
\left\langle k_{r}^{2} \right\rangle
 =
\frac{1}{N}
\int{{{\nu }_{{{\mathbf{k}}_{r}}}}k_{r}^{2}d{{\mathbf{k}}_{r}}}
\\
=
\frac{{{g}^{*}}q}{4N}\frac{T}{{{m}^{*}}{{\Omega }^{2}}}\int\limits_{0}^{\infty }
{
\ln {{\left[ 1-\exp \left( -\frac{{{\hbar }^{2}}k_{r}^{2}}{2{{m}^{*}}T} \right) \right]}^{-1}}k_{r}^{2}
d\left( k_{r}^{2} \right)
}.
\end{gathered}\end{split}\end{equation}

\no
After some hackneyed algebra, one can obtain an expression for the mean value of the transverse wave vector squared,

\begin{equation}\begin{split}\begin{gathered}
\label{mean k^2}
\left\langle k_{r}^{2} \right\rangle =\zeta \left( 3 \right){{g}^{*}}q\frac{{{m}^{*}}{{T}^{3}}}{{{N}}{{\hbar }^{4}}{{\Omega }^{2}}},
\end{gathered}\end{split}\end{equation}

\noindent
where $\zeta \left( 3 \right)\approx 1.2$ refers to Riemann zeta function. By substituting Eq.\eqref{mean k^2} into Eq.\eqref{spectral width def}, one obtains the blur of the condensed light signal:

\begin{equation}\begin{split}\begin{gathered}
\label{spectral width}
\hbar \left\langle \omega -{{\omega }_{0}} \right\rangle =\frac{\zeta \left( 3 \right){{g}^{*}}q}{2{{N}}}\frac{{{T}^{3}}}{{{\left( \hbar \Omega  \right)}^{2}}}.
\end{gathered}\end{split}\end{equation}

\no
Taking into account Eq.\eqref{above cond particles}, formula \eqref{spectral width} can be also written in the form:

\begin{equation}\begin{split}\begin{gathered}
\label{spectral width 2}
\hbar \left\langle \omega -{{\omega }_{0}} \right\rangle =6{{\pi }^{-2}}\zeta \left( 3 \right)\left( \frac{{{N}_{{{\mathbf{k}}_{r}}\ne 0}}}{{{N}}} \right)T.
\end{gathered}\end{split}\end{equation}

\no
Expression \eqref{spectral width 2} shows that for the fixed relative amount of condensed photons $N_{\mathbf{k}_{r} \ne 0} / N$ , the  blur caused by thermal fluctuations of non-condensed photons is linear on the system's temperature $T$.

One can also calculate the higher moments of spectral density ${{\nu }_{{{\mathbf{k}}_{r}}}}$ in the center of the condensed cloud, $r \approx 0$.
Similar to the expression \eqref{mean k^2 def}, the quantity ${{\hbar }^{2}}\left\langle {{\left( \omega -{{\omega }_{0}} \right)}^{2}} \right\rangle $, which is the variance in mathematical point of view, can be found as:

\begin{equation}\begin{split}\begin{gathered}
\label{variance}
{{\hbar }^{2}}\left\langle {{\left( \omega -{{\omega }_{0}} \right)}^{2}} \right\rangle 
=
\left( \frac{\hbar^{3}}{2 m^{*}} \right)^2
\left\langle k_{r}^{4} \right\rangle
\\
=
\frac{{{\pi }^{4}}}{90}\frac{{{g}^{*}}q}{{{N}}}\frac{{{T}^{4}}}{{{\left( \hbar \Omega  \right)}^{2}}}.
\end{gathered}\end{split}\end{equation}

\no
Therefore,  one can estimate  the standard deviation as 

\begin{equation}\begin{split}\begin{gathered}
\label{deviation}
\hbar \sqrt{\left\langle {{\left( \omega -{{\omega }_{0}} \right)}^{2}} \right\rangle }
\sim
{{\left( \frac{{{N}_{{{\mathbf{k}}_{r}}\ne 0}}}{{{N}}} \right)}^{1/2}}T.
\end{gathered}\end{split}\end{equation}

According to expressions \eqref{spectral width 2} -\eqref{deviation}, the effects caused by the thermal blur are suppressed when the number of condensed photons $N_{\mathbf{k}_{r}=0}$ increases. As a consequence, the observed spectral peak becomes sharper if the amount of condensed photons grows.

To compare with the current experimental results on Bose-Einstein condensation of photons in a microcavity [see Refs.\cite{thermalization},\cite{Klaers}], we make here the following numerical estimate. For the parameters of the system, similar to the parameters of experimental setup, and the  relative amount of condensed photons  $N_{\mathbf{k}_{r} \ne 0} / N \approx 50 \% $, formula \eqref{spectral width 2} gives us the value $\hbar \left\langle \omega -{{\omega }_{0}} \right\rangle \sim {{10}^{-2}} \ eV$. 
Note that a similar thermal blur is observed in experiment  \cite{Klaers}.

We now proceed to the description of spatial parameters of observed phenomenon. By analogy with a trapped BEC of ultracold alkali gases, we shall call a spatially localized thickening of condensed particles as a "cloud". The average spatial extent of a cloud of the condensed photons can be estimated as:

\begin{equation}\begin{split}\begin{gathered}
\label{cloud radius}
\left\langle r \right\rangle 
=\frac{1 }{{{N}}}\int\limits_{0}^{\infty }{r\left[ {{n}_{{{\mathbf{k}}_{r}}=0}}(r)+{{n}_{{{\mathbf{k}}_{r}}\ne 0}}(r) \right]} \, d\left( \pi {{r}^{2}} \right).
\end{gathered}\end{split}\end{equation}

\no
Recall that the exact spatial dependence of the density of condensed photons  ${{n}_{{{\mathbf{k}}_{r}}=0}}(r)$ is unknown.  Therefore one cannot  perform explicitly the averaging in \eqref{cloud radius}. 
So we  go the following way. Apparently on can rewrite the expression \eqref{cloud radius} in the form:

\begin{equation}\begin{split}\begin{gathered}
\label{cloud extent}
\left\langle r \right\rangle 
=
\frac{{{N}_{{{\mathbf{k}}_{r}}=0}}}{{{N}}}{{\left\langle r \right\rangle }_{T=0}}+\frac{{{N}_{{{\mathbf{k}}_{r}} \ne 0}}}{{{N}}}{{\left\langle r \right\rangle }_{T\ne 0}}.
\end{gathered}\end{split}\end{equation}

\noindent
where ${{\left\langle r \right\rangle }_{T=0}}$ is a typical radius of the condensed cloud if the thermal blur is absent, and  ${{\left\langle r \right\rangle }_{T \ne 0}}$ is the contribution of non-condensed photons to the smearing of the condensed cloud.
For the first  quantity, the following estimate in order of magnitude is legitimate:

\begin{equation}\begin{split}\begin{gathered}
\label{spatial extent 1}
{{\left\langle r \right\rangle }_{T=0}}\sim a,
\end{gathered}\end{split}\end{equation}

\noindent
where $a$ is a typical spatial extent of a trapped BEC with negligibly small interactions between composing bosons [for details see for example Ref.\cite{Pethick Smith}],

\begin{equation}\begin{split}\begin{gathered}
\label{oscillator length}
a=\sqrt{\frac{\hbar }{{{m}^{*}}\Omega }}.
\end{gathered}\end{split}\end{equation}

\no
The quantity  ${{\left\langle r \right\rangle }_{T \ne 0}}$, appearing in formula \eqref{cloud extent}, depends on the spatial distribution of non-condensed photons ${{n}_{{{\mathbf{k}}_{r}} \ne 0}}(r)$ which is known even beneath the BEC transition point. It can be easily obtained from the Eq.\eqref{2D density} taking into account expression \eqref{mu*}.
Therefore, taking into account Eq.\eqref{above cond particles}, the "thermal" contribution to the smearing of a cloud can be calculated as:

\begin{equation}\begin{split}\begin{gathered}
\label{spatial extent 2}
{{\left\langle r \right\rangle }_{T\ne 0}}=
\frac{3\sqrt{2}}{{{\pi }^{3/2}}}\, \zeta (5/2 ) \,
\sqrt{\frac{T}{{{m}^{*}}{{\Omega }^{2}}}}.
\end{gathered}\end{split}\end{equation}

\no
Note that the quantity $\sqrt{T/{{m}^{*}}{{\Omega }^{2}}}$ defines a typical distance where the effective "potential" energy [the third term on the right-hand side of the expression  \eqref{E*}]  becomes in the order of the system's temperature $T$. 
We also note here that the ratio between the two spatial parameters depends only on the amount of non-condensed photons:

\[\frac{{{\left\langle r \right\rangle }_{T\ne 0}}}{{{\left\langle r \right\rangle }_{T= 0}}}\sim \sqrt{\frac{T}{\hbar \Omega }}\sim \sqrt[4]{{{N}_{{{\mathbf{k}}_{r}}\ne 0}}} \, .\]

\no
Therefore, according to the formulas \eqref{cloud extent}-\eqref{spatial extent 2}, by lowering the temperature of the system $T$, or increasing the relative amount of condensed photons for the fixed temperature, the radius of the condensed cloud is reduced.

It is appropriate to make here the following numerical estimates.  Using expressions \eqref{oscillator length}-\eqref{spatial extent 2} for the parameters of the system, similar to the parameters of experimental setup in Ref.\cite{Klaers}, one obtains numerical values $a \sim 10^{-5} m$ and ${{\left\langle r \right\rangle }_{T\ne 0}} \approx 10^{-5} \, m$.
Therefore, one can conclude that both of the spatial extents ${{\left\langle r \right\rangle }_{T=0}}$, ${{\left\langle r \right\rangle }_{T \ne 0}}$ give approximately equal contributions and thus noone of them can be neglected in the present system.
Note also that our calculations are in a good agreement with experimental observations: in the experiments \cite{Klaers}, the spatial extent of the cloud in an intermediate case (i.e. when there is a condensate, but not all of the photons have been condensed) is approximately $4 \cdot 10^{-5} \  m$.

Finally, we  calculate thermodynamic properties of the photon gas with condensate. First, it is interesting to calculate the total energy of the photons after they have been thermalized with medium inside cavity. The easiest way to do it is to add the energy of condensed photons to the energy of photons with non-zero transverse momentum:

\begin{equation}\begin{split}\begin{gathered}
\label{full energy def}
E=
{{N}_{{{\mathbf{k}}_{r}}=0}} \, \hbar \omega_{0}+
\frac{{{g}^{*}}}{{{\left( 2\pi  \right)}^{3}}}\sum\limits_{{{k}_{z}}}{\int\limits_{z}{dz}}
\int\limits_{0}^{\infty }{ d\left(\pi k_{r}^{2} \right)}
\\
\times
\int\limits_{0}^{\infty }{ d\left(\pi {{r}^{2}} \right)}\frac{\hbar {{\omega }_{0}}+{{\hbar }^{2}}k_{r}^{2}/2{{m}^{*}}+{{m}^{*}}{{\Omega }^{2}}{{r}^{2}}/2}{\exp \left[ \frac{{{\hbar }^{2}}k_{r}^{2}}{2{{m}^{*}}T}+\frac{{{m}^{*}}{{\Omega }^{2}}{{r}^{2}}}{2T} \right]-1}.
\end{gathered}\end{split}\end{equation}

\no
Calculating the integrals in \eqref{full energy def}, one can obtain the following temperature dependence of the light total energy beneath the BEC transition point:

\begin{equation}\begin{split}\begin{gathered}
\label{full energy}
E=N \hbar {{\omega }_{0}}+{\zeta \left( 3 \right){g}^{*}}q
\frac{{{T}^{3}}}{{{\left( \hbar \Omega  \right)}^{2}}}.
\end{gathered}\end{split}\end{equation}

\no
We emphasize here that photons in the BEC state possess the total energy of amount:

\begin{equation}\begin{split}\begin{gathered}
{{E}_{{{\mathbf{k}}_{r}}=0}}={{N}_{{{\mathbf{k}}_{r}}=0}}\,\hbar {{\omega }_{0}}.
\end{gathered}\end{split}\end{equation}

\no
We can also introduce the heat capacity of the photonic system in a conventional way,  $C \equiv dE/dT$.  One should keep in mind that the number of the photons thermalized with medium in the general case depends on the temperature of the medium, $N=N(T)$. If this dependence is relatively weak, $dN/dT \approx 0$, 
 we can define the heat capacity of the photonic gas, thermalized with in-cavity medium  beneath the phase transition point as follows:

\begin{equation}\begin{split}\begin{gathered}
\label{heat capacity}
C \approx 3\zeta \left( 3 \right){{g}^{*}}q\frac{{{T}^{2}}}{{{\left( \hbar \Omega  \right)}^{2}}}.
\end{gathered}\end{split}\end{equation}

\no
Closer to the critical point one should also take into account the discontinuity in specific heat caused by the singularity in the behavior of the chemical potential of photons (see e.g. Ref.\cite{Pethick Smith})

\[\Delta C \propto N{\left. {\frac{{\partial {\mu ^*}\left( T \right)}}{{\partial T}}} \right|_{T = {T_{c + }}}}\]

\no 
where $T_{c + }=T_c+0$ denotes the upper side of the transition point $T_c$. In the general case, the dependence $\mu^{*} (T) $ for $T>T_c$ can be calculated numerically or estimated analytically in the symmetric phase (see e.g. Ref.\cite{Kruchkov}).

The expression \eqref{full energy} can be also formulated  in a more traditional appearance by using the definition for the total number of non-condensed photons \eqref{above cond particles} :

\begin{equation}\begin{split}\begin{gathered}
E=N \, \hbar {{\omega }_{0}}+\frac{12 \zeta (3) }{{{\pi }^{2}}}{{N}_{{{\mathbf{k}}_{r}}\ne 0}}\, T.
\end{gathered}\end{split}\end{equation}

\no
Note that $12 \zeta (3) / \pi^{2} \approx 1.46$  is not far from the factor $\frac{3}{2}$ for the energy of an ideal three-dimensional gas of classical particles with the fixed total number ${{N}_{{{\mathbf{k}}_{r}}\ne 0}}$.

One can introduce the quantity $\eta$ showing the ratio of the energy of condensed photons to the total light energy in the system under study:

\begin{equation}\begin{split}\begin{gathered}
\label{eta}
\eta =\frac{{{E}_{{{\mathbf{k}}_{r}}=0}}}{E}
\\
=
\frac{{{N}_{{{\mathbf{k}}_{r}}=0}}}{N}{{\left[ 1+\frac{12\zeta (3)}{{{\pi }^{2}}}\left( 1-\frac{{{N}_{{{\mathbf{k}}_{r}}=0}}}{N} \right)\frac{T}{\hbar {{\omega }_{0}}} \right]}^{-1}}.
\end{gathered}\end{split}\end{equation}

\no
Obviously, there are conditions when $\eta \approx 1$ and therefore all the in-cavity light radiation energy will be accumulated in the Bose-condensed state of light. According to Eq.\eqref{eta}, one can control and manipulate the efficiency of the energy conversion $\eta$ to maximize the amount of light energy stored in the monoenergetic BEC state. It could be done by varying both the geometrical and the thermodynamical parameters of a cavity with a medium.

\section{Conclusion}

The purpose of the present paper was to examine the influence of intracavity medium on the parameters of the Bose-Einstein condensation of photons, and also to derive consequently all the statistical properties, which can explain and describe the observable quantities numerically. In the assumption of gaseous (i.e. weakly interacting) medium, we show that the critical number of photons to create the  BEC state, as well as all other statistical characteristics, indeed depend on the description parameters of matter. In this sense we eliminate the disagreement over why the thermostatic medium which is so necessary to thermalize photons, has not been involved into the critical parameters of the BEC transition. Talking here about a gaseous medium we actually do not restrict ourselves to the case of an atomic gas: In point of fact, one can consider an ensemble of organic dye molecules as a weakly interacting gas in a solvent (see also \cite{Klaers}).

One of the basic assumptions in our model, and in the models of other studies (see Refs.\cite{Klaers,Kruchkov,Sobyanin2,Leeuw}), was a so-called two-level model of gaseous medium. In this model all the structural units of a gaseous medium (atoms or molecules) can be only either in the ground state or in the first excited state, and the transition between these two states is caused uniquely by an absorption or re-emission of a photon. Nevertheless, it is not a fundamental restriction, and our theory, of course, can be generalized to the case with an arbitrary number of excited states. However, in fact, for the current state of experiments there is no need to do it: Apparently, the number of elementary excitations in the system under study is relatively small, and the probability to excite the higher molecular states is vanishing. This fact was also used in Ref. \cite{Kruchkov} to calculate the properties of a possible three-dimensional photon condensate in ideal gases (including some types of plasma). 
Of course, in a more general case,  the degeneracy of a photon state may depend on its energy. From the general considerations it is clear that the numerous and repeated processes of absorption and reemission which thermalize light radiation inside cavity and also the boundary conditions on cavity walls should influence the photon degeneracy in the system under study [see also \cite{Kruchkov}]. This is a separate and nontrivial problem that is beyond the scope of our paper. However, we recall that in the current experiments  \cite{Klaers} all the pumped photons have approximately the same energy, which is very close to $\hbar \omega_{0}$. Therefore one can actually introduce an effective degeneracy $g^{*}$ for all the thermalized  photons. For numerical estimates in the present paper we used $g^{*} \approx 2 $ (see also \cite{Kruchkov}). This assumption is possible because the average thermal fluctuation of the photon energy is relatively small. 
One should also keep in mind that the neighborhood of the critical point should be treated more carefully (see e.g. \cite{Kocharovsky,Sobyanin2,Ketterle,Slyusarenko}). 
However, the mentioned limitations are not principal and the present theory can be expanded and improved for a better accuracy.

Despite the use of a relatively simple model, we have been able to represent the crucial characteristics of photon BEC and therefore explain the experiments. It is worth emphasizing the very good correlation between the physical quantities  estimated in the present paper (such as the  thermal blur width and the special extent of condensed cloud of photons) and those ones observed in experiment  \cite{Klaers}.  Such an excellent agreement one more time justifies the used simplifications of our model. 

In conclusion, we also want to mark the importance of the direct demonstration of the fact that light energy can be accumulated in the Bose-Einstein-condensed state.
We also showed, that the energy conversion factor $\eta$ can be maximized by manipulating the cavity parameters, the temperature of the system or by changing the amount of photons. This gives a good perspective to implement the phenomenon of light condensation as the working principle of different solar cells. In this sense the question rises as to whether it is possible to condense more than one light mode in the same cavity. The possibility of a multi-level photon condensation is still an open question.

\section{Appendix}

Despite the fact that the system described in the present paper reveals effectively two-dimensional properties, it nevertheless  remains purely three-dimensional. Therefore one should carefully approach to the derivation of the effective two-dimensional (2D) density of photons and calculation of all the average quantities.

The reduction to the 2D distribution function can be done by the procedure of integration over longitudinal coordinate $z$ with subsequent summing over all the possible longitudinal wavevectors $k_{z}$.
Current experiments are conducted just for the one mode of electromagnetic field inside a cavity (for example, it is $q=7$ or $q=11$ in the experiments of Ref.\cite{Klaers}).
Consequently, the mentioned procedure comes to multiplication on the constant factor,

\[\sum\limits_{{{k}_{z}}}{\int\limits_{z}{dz \left( ... \right)}=}\pi q\left( ... \right)\]

This procedure allows us to calculate correctly statistical averages, for example, the total light energy inside a cavity.

\end{document}